\documentclass[seceq]{ptptex}

\usepackage{graphicx}% Include figure files
\usepackage{wrapft}

\title{%        %You can use \\ for explicit line-break
Testing gravitational physics with superconducting gravimeters%
}

\author{%       %Use \scshape  for the family name
Sachie \textsc{Shiomi}%
}

\inst{%     %Affiliation, neglected when [addenda] or [errata]
Space Geodesy Laboratory, Department of Civil Engineering,\\
National Chiao Tung University, 1001 Ta-Hsueh Rd., Hsinchu, Taiwan
300 R.O.C. }

\abst{%         %this abstract is neglected when [addenda] or [errata]

Superconducting gravimeters are the most sensitive instruments to
measure surface gravity changes at low frequencies. Currently, about
twenty five superconducting gravimeters are operating in the world
and their global network has been developed. We investigate possible
applications of the superconducting gravimeters to tests of
gravitational physics. Previous experimental searches for spacial
anisotropies in the gravitational constant $G$ and for gravitational
waves, performed with gravimeters in 1960's to 1970's, can be
improved by applications of the current superconducting gravimeters.
Also, we describe other proposed applications of testing the
universality of free-fall and searching for composition-dependent
dilatonic waves, and discuss future works necessary for these
geophysical tests.}

\begin{document}

\maketitle

\section{Introduction}

Superconducting gravimeters are the most sensitive instruments of
measuring gravity at low frequencies. They have been used for tests
of gravitational physics and geophysical studies
(Ref.~\citen{Goodkind1999} and references therein). The global
network of superconducting gravimeters, the Global Geodynamics
Project (GGP) network \cite{GGP}, has been developed since 1997.
Currently, about twenty-five superconducting gravimeters join the
GGP network (see section \ref{st:GGP} for the distribution of the
GGP stations). The GGP network allows us to study gravity signals in
global nature with increased sensitivity. It has been successfully
used for geophysical studies \cite{Hinderer2004}. Its applications
to gravitational physics were suggested \cite{Goodkind1999,
Shiomi2006, Shiomi2007}, but they have not been studied in detail
yet. In this paper, we will focus on possible applications of the
global network to gravitational physics.

Brief descriptions of the instrument and the GGP network are given
in sections \ref{st:SG} and \ref{st:GGP}, respectively. We will see
previous applications of superconducting gravimeters in
gravitational physics and discuss possible future applications of
superconducting gravimeters and the GGP network in gravitational
physics in section \ref{st:Applications}.

\section{Superconducting gravimeters}\label{st:SG}
The superconducting gravimeters were developed by Goodkind and
Prothero at the University of California, San Diego (UCSD), in the
late 1960's \cite{Prothero1968, Goodkind1999}. The fundamental
design described in their first report\cite{Prothero1968} has not
been changed, but its performance has been improved since then at
UCSD and GWR Instruments \cite{GWR}. Currently, commercial
superconducting gravimeters are available at GWR Instruments.

\begin{figure}
\includegraphics[width=\linewidth]{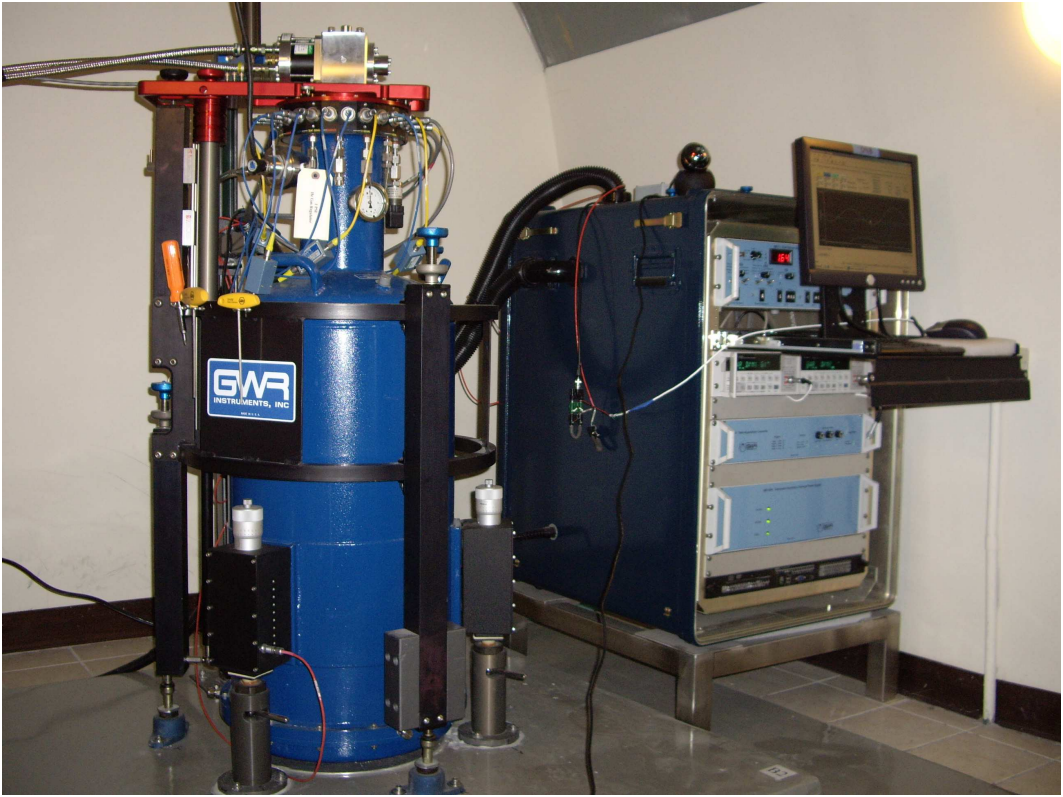}
\caption{A photograph of the superconducting gravimeter (No. 48) and
its data acquisition system, installed at LOGG in Hsinchu Taiwan.
The sensing unit (Fig. \ref{fig:SG-GSU}) is placed in the liquid
helium dewar (the blue tank in this photograph) and operates at
liquid helium temperatures ($\sim$ 4.2 K).} \label{fig:SG-photo}
\end{figure}

\begin{center}
\begin{figure}
\includegraphics[height=10 cm]{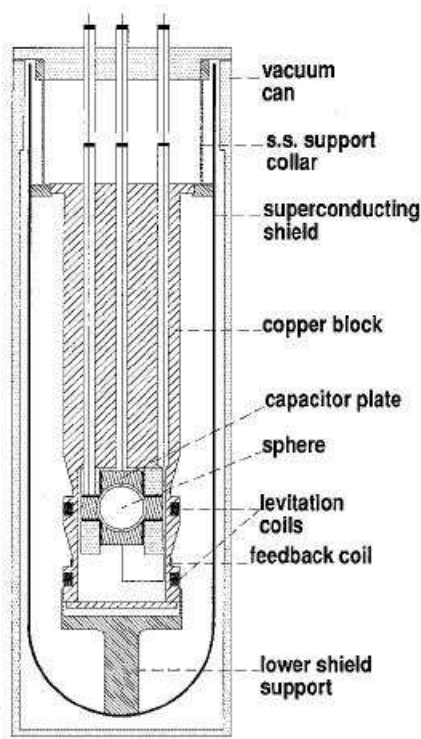}
\caption{A schematic cross-section of the sensing unit of a
superconducting gravimeter (quoted from Ref.~\citen{Goodkind1999}).}
\label{fig:SG-GSU}
\end{figure}
\end{center}

Two new superconducting gravimeters have been installed at the
Laboratory of Geodesy and Geodynamics (LOGG) in Hsinchu Taiwan
(24.8$^{\circ}$N, 121$^{\circ}$E) in March 2006. They were
manufactured by GWR Instruments. A photograph of one of the
superconducting gravimeters (No. 48) is shown in Fig.
\ref{fig:SG-photo}. A schematic view of the sensing unit of the
superconducting gravimeter is shown in Fig. \ref{fig:SG-GSU} (quoted
from Ref.~\citen{Goodkind1999}).

In a superconducting gravimeter, instead of the spring used in a
mechanical gravimeter, its proof mass (a superconducting sphere) is
levitated by magnetic fields, induced in superconducting levitation
coils (see Fig. \ref{fig:SG-GSU}). The proof mass is about 2.5 cm
(one inch) in diameter and its weight is between 4 and 8 g
\cite{Goodkind1999}. By adjusting the currents of the levitation
coils, the magnetic stiffness (spring constant) can be tuned to be
nearly zero. The motion of the proof mass, in response to changes in
ambient gravity, is monitored by capacitive sensors that surrounds
the proof mass. In operation, the proof mass is kept at the same
position through a feed-back system. Because of the stability in the
super currents in the levitation coils and the smallness in
stiffness, superconducting gravimeters provide stable and sensitive
gravity measurements. The sensitivity of a superconducting
gravimeter, installed at a quiet site, is better than $\sim$ 1 $n$
gal or 10$^{-11}$ m s$^{-2}$ for a year-long measurement at various
frequencies and its stability is better than a few $\mu$ gal
(10$^{-8}$ m s$^{-2}$) per year for resent instruments. A more
detailed description of superconducting gravimeters is given in
Ref.~\citen{Goodkind1999}.

\section{The GGP network}\label{st:GGP}

The GGP network \cite{GGP} has been developed since 1997 to study
geophysical signals in global nature, for example, the inner core
oscillations, polar motion and wobbles \cite{Crossley2004}. The map
of the GGP stations is given in Fig. \ref{fig:GGP}. Currently about
twenty five stations join the GGP network. From Fig. \ref{fig:GGP},
one can see that the GGP stations are widely distributed, from north
to south and east to west on the globe.

\begin{figure}
\includegraphics[width=\linewidth]{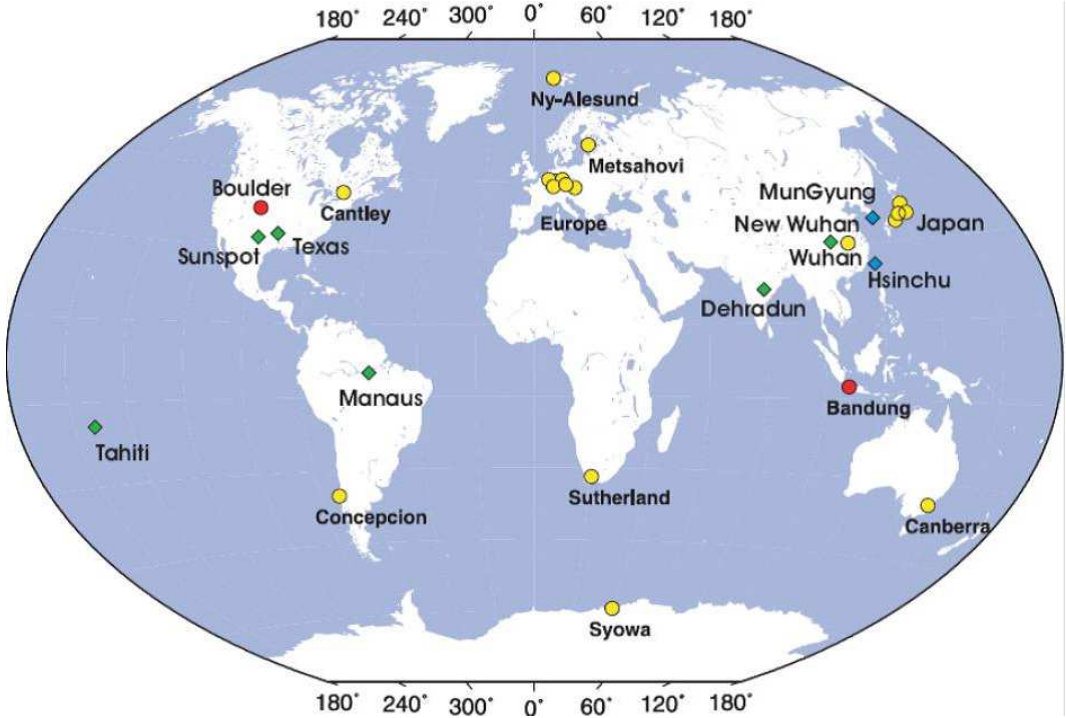}
\caption{The map of the GGP stations, quoted from
Ref.~\citen{Crossley2007}. Yellow circles indicate the stations that
are currently operating. Red circles indicate the stations that
recently stopped operation. Green diamonds indicate the stations
that are going to install superconducting gravimeters in the near
future. Blue diamonds indicate the stations that newly installed
superconducting gravimeters.}
\end{figure}\label{fig:GGP}

\section{Applications in gravitational
physics}\label{st:Applications}

Superconducting gravimeters have been applied to tests of
gravitational physics since 1970's. One of the earliest works in the
context is the search for anisotropies in locally measured values of
the gravitational constant $G$ \cite{Warburdon1976}, which are
predicted by some gravitation theories \cite{Will1971,
Nordtvedt1972, Will1973}. It is pointed out that this search can be
improved with the use of longer records of gravity data from
multi-stations \cite{Goodkind1999}. We will discuss the possible
improvement in section \ref{sst:anisotopies}.

Other experiments, which have been performed using superconducting
gravimeters, are testing the inverse-square law in a laboratory
scale (0.4 to 1.4 m)\cite{Goodkind1993} and in the geophysical
window (10 to 10$^3$ m) \cite{Achilli1997, Baldi2001}, a
determination of the gravitational constant $G$ \cite{Baldi2005},
and searching for gravitational waves using the Earth as the
receiver \cite{Tuman1971,Tuman1973}. We will describe the
gravitational-wave search in more detail in section
\ref{sst:GWsearch}.

As for future experiments, we will briefly describe two proposals:
testing the universality of free-fall \cite{Shiomi2006} (section
\ref{sst:VEP}) and searching for composition-dependent dilatonic
waves (section \ref{sst:DW}).

\subsection{Search for spacial anisotropies in $G$}\label{sst:anisotopies}
Unlike General Relativity, some theories of
gravitation predict spacial anisotropies in locally measured values
of the gravitational constant $G$ (References~\citen{Will1971,
Nordtvedt1972, Will1973} and references therein). Among these
theories, some allow the existence of preferred frames in the
universe, and such anisotropies in laboratory-measured values of $G$
arise from the translation and rotation of the Earth relative to the
assumed preferred frames (preferred-frame effects). In another type
of theories, anisotropies in $G$ are caused by a nearby gravitating
body, such as the Galaxy (the Galaxy induced anisotropy or
preferred-location effects). These anisotropies lead to anomalous
tidal effects, which can be searched for with gravimeters.

Will has examined Earth-tide data obtained from mechanical
gravimeters and found that they agree with Newtonian predictions
within two percent \cite{Will1971}. This indicates that the
magnitude of the anomalous tidal effects should be less than
10$^{-9}g$, where $g \approx 9.8$ m s$^{-2}$ is the Earth's surface
gravitational field. With this experimental limit, he obtained upper
limits on the preferred-frame effects and the preferred-location
effects, which are parameterized by $\alpha_2$ and $\xi$ in the
parameterized post-Newtonian (PPN) formalism\footnote{see
Ref.~\citen{Will2006} and references therein for detailed
descriptions of the PPN formalism.}, respectively: $\alpha_2 < 3
\times 10^{-2}$ and $\xi < 10^{-2}$ \cite{Will1971}.

Warburdon and Goodkind \cite{Warburdon1976} searched for such
anomalous tidal effects in their gravity data of superconducting
gravimeter and placed more stringent upper limits: $\alpha_2 < 4
\times 10^{-4}$ and $\xi < 10^{-3}$. This upper limit on $\xi$ is
currently the most stringent constraint on the PPN parameter (see
Table 4 on p. 43 in Ref.~\citen{Will2006} for current limits on the
PPN parameters).

Later, it is shown that $\alpha_2$ can be constrained to be order of
10$^{-7}$ from the close alignment of the Sun's spin axis with the
solar system's planetary angular momentum after 5 billion yr
\cite{Nordtvedt1987}. Also, it is shown that $\alpha_2$ should be
determined to a few parts in 10$^{-5}$ using Lunar Laser Ranging
(LLR) data \cite{Nordtvedt1996}, and its preliminary estimate is
given as $\alpha_2 = (2 \pm 2) \times 10^{-5}$ in
Ref.~\citen{Muller2007}.

With the use of longer records of gravity data from the GGP stations
and the improved knowledge of geophysical and environmental
disturbances, we could improve the estimates of the upper limits
placed by Warburdon and Goodkind. From recent observations, it is
shown that the noise level of a superconducting gravimeter located
at a quiet site is about a few ngal (10$^{-12}g$) \cite{Rosat2004}
at the signal frequencies of the preferred-frame and
preferred-location effects. This is about three orders of magnitude
improvements in gravity measurements in comparison with the
experimental data used by Will in early 1970's. With this current
noise level, we could obtain upper limits on both of $\alpha_2$ and
$\xi$ in order of 10$^{-5}$, which is comparable with the expected
sensitivity from the LLR data. Further analyses are necessary to
obtain accurate estimates.

\subsection{Search for gravitational waves}\label{sst:GWsearch}
In the early stage of experimental studies on detecting
gravitational waves, a pioneer of gravitational-wave research,
Weber, proposed to search for normal modes of the Earth, exited by
incident gravitational waves \cite{Weber1960}. The first upper
limits on the flux of gravitational waves were placed by checking
the excitation of the normal modes, using seismological data
\cite{Forward1961} and a mechanical gravimeter \cite{Weber1967}, in
1960's. In the early 1970's, possible excitations of the normal
modes were observed with a superconducting gravimeter
\cite{Tuman1971, Tuman1973}, but the results have not been confirmed
by following experiments.

This approach of searching for gravitational waves is sensitive at
low frequencies (about 0.3 mHz or 54 min in period for the $_0S_2$
mode); it allows to investigate lower frequencies than other
ground-based gravitational-wave detectors (i.e. laser
interferometers and resonant-mass detectors), whose sensitive
frequency ranges are about 10 to 1000 Hz \cite{GWIC}.
Superconducting gravimeters are suitable for the normal-mode method
because of their good sensitivity at the low-frequency range.
Superconducting gravimeters were improved from the one used for the
gravitational-wave search in early 1970's \cite{Tuman1971,Tuman1973}
and the GGP network is now available. Our knowledge of low-frequency
normal modes and the Earth model has been being improved in
geophysics \cite{Rosat2005,Roult2006}. Therefore, it might be
interesting to reinvestigate the normal-mode method using current
superconducting gravimeters and the GGP network.

In the frequency range of the normal-mode method, there are
stringent upper limits on the cosmological stochastic gravitational
waves by the Nucleosynthesis and recent measurements of the cosmic
microwave background: $\Omega_{gw}h^2_{100} \lesssim 10^{-5}$
\cite{Cyburt2005,Smith2006}. As for astrophysical stochastic
gravitational waves, the Doppler tracking of the Cassini spacecraft
has placed an upper limit: $\Omega_{gw}h^2_{100} \lesssim 0.1$ at
$\sim$ 0.3 mHz \cite{Armstrong2003}. To provide a significant
contribution to the filed, we have to achieve a comparable
sensitivity with the Doppler tracking result.

The normal-mode method has fundamental difficulties to separate the
expected signals of gravitational waves from seismic noise and
geophysical effects, such as excitations of the normal modes by
silent earthquakes \cite{Beroza1990}. However, with the use of the
GGP network, we could identify some localized disturbing effects and
remove them from the gravity data. Further investigation is
necessary.

Also, scalar gravitational waves, predicted by scalar-tensor
theories of gravitation (e.g. Brans-Dicke's theory), can be searched
for by checking excitations of spherically symmetric modes of the
Earth, during quiet periods. Some efforts of studying the
twenty-minute breathing mode, $_0S_0$, have been done by Block,
Weiss and Dicke in an early stage of gravitational-wave research
\cite{Weber1964}. This mode can now be studied with improved
sensitivity by using current superconducting gravimeters and the GGP
network.

%With improved knowledge of the Earth's interior, we could model and
%remove the unwanted effects with a better precision in the future.

\subsection{Test of the universality of free-fall} \label{sst:VEP}

The Earth's inner core is weakly coupled to the rest part of the
Earth by mainly gravitational forces. If there were a violation of
the universality of free-fall, because of their different chemical
compositions and/or of different mass fractions of binding energies,
the inner core and the rest part of the Earth would fall at
different rates towards the Sun and other sources of gravitational
fields \cite{Shiomi2006}. The differential acceleration would result
in surface-gravity effects, which can be searched for using
superconducting gravimeters. Based on a simple Earth model, it is
shown that the universality can be tested to a level of 10$^9$ with
a superconducting gravimeter \cite{Shiomi2006}. To be comparable
with current best limits on tests of the universality
\cite{Fischbach1999}, the sensitivity has to be improved by more
than three orders of magnitude. There are several possibilities to
improve the sensitivity. One way is to apply advanced data analysis
methods to extract weak signals. According to a non-linear
damped-harmonic analysis method used in geophysical studies, it is
possible to improve the sensitivity by a factor of $\sim$ 10
\cite{Rosat2007}. Another way may be to carry out coincidence
measurements with two superconducting gravimeters located ideally
opposite sides of the Earth near the equator. If there were a
violation towards the Sun, the expected magnitude of the violation
signal at the two superconducting gravimeters is the same but the
sign should be opposite. By combining such coincidence signals, we
could double the magnitude of the expected signals and the
sensitivity would be improved by a factor of 2.

Because of the inclination of the Earth's rotation axis, the maximum
violation signals towards the Sun can be expected at observatories
located on the equator in Spring and Autumnal equinox points, and on
Tropic of Cancer or Capricorn in Summer and Winter solstices
\cite{Shiomi2006}. Our site in Taiwan is one of the ideal locations
for this approach in Summer and Winter. If the noise level of data
from our site is high, it might be better to use data from low noise
sites considering the degrees of signal compensation depending on
the latitude and longitude of the sites. Further studies are
necessary to figure out the optimal schemes for global observations
and noise reduction.

A more detailed description of this geophysical test of the
universality in given in Ref.~\citen{Shiomi2006}.

\subsection{Search for dilatonic waves}\label{sst:DW}

Composition-dependent dilatonic waves are predicted by unified
theories of strings \cite{Gasperini2003}. When such dilatonic waves
pass the Earth, because of the difference in dilatonic charge
(namely, the difference in the chemical compositions) between the
Earth's inner core and the rest part of the Earth, there would be
relative motions between them \cite{Shiomi2007}. Such relative
motions would result in surface gravity changes, which can be
searched for by superconducting gravimeters. This method has its
best sensitivity at the resonant frequency of the translational
motions of the inner core:$\sim$ 7 $\times$ 10$^{-5}$ Hz, which is
lower than the sensitive frequencies of previous proposals using
gravitational-wave detectors: $\sim$ 10 to 1000 Hz. Using available
results of surface-gravity measurements with superconducting
gravimeters and assuming a simple Earth model, preliminary upper
limits on the energy density of dilatonic waves can be obtained at
the low frequency. However, the results are currently limited by the
uncertainty in the Earth model. A more detailed description of this
method and the preliminary results are given in
Ref.~\citen{Shiomi2007}.

\section{Summary and discussions}

We have discussed the following geophysical tests of gravitational
physics in the previous sections: searching for preferred-frame and
preferred-location effects (section \ref{sst:anisotopies}),
searching for gravitational waves and scalar gravitational waves
(section \ref{sst:GWsearch}), testing the universality of free-fall
(section \ref{sst:VEP}) and searching for composition-dependent
dilatonic waves (section \ref{sst:DW}). These discussed applications
are summarized in Table \ref{table:summary}.

From the third column of the table, one can see that the frequencies
of the searched for effects are low: $\sim$ $10^{-5}$ to $10^{-3}$
Hz. In the low frequency range, superconducting gravimeters are the
most sensitive instruments.

\begin{table}
\caption{Summary of the geophysical tests of gravitational physics,
discussed in section \ref{st:Applications}. The second and third
columns indicate the expected dominant phenomena and their periods,
respectively.}\label{table:summary}
\begin{center}
\begin{tabular}{|l|l|l|}
  \hline
  Searched for signals & Expected phenomena on the Earth & Periods
  \\ \hline
  Preferred-frame effects ($\alpha_2$) & Anomalous tides & 12 hours \\
  Preferred-location effects ($\xi$) & Anomalous tides & 12 hours \\ \hline
  Gravitational waves & Excitation of the $_0S_2$ mode & 54 minutes \\
  Scalar gravitational waves&  Excitation of the $_0S_0$ mode & 20.5 minutes
  \\ \hline
  Violation of the universality of free-fall & Translational motions of the inner core & $\sim$ 24 hours \\
  Composition-dependent dilatonic waves & Excitation of translational motions  & 4-6 hours \\
                                        &of the inner core& \\
  \hline
\end{tabular}
\end{center}
\end{table}

Those searched for effects are all thought to be very small. In
order to have significant contributions to the field of
gravitational physics, it is essential to improve the sensitivity.
Key researches and developments to improve the sensitivity may be
(1) developing data analysis methods to extract weak signals, (2)
figuring out the optimum use of the global data, (3) carrying out
coincidence measurements and (4) improving the Earth model.

As mentioned earlier, data analysis methods to extract weak signals
have been being studied in geophysics \cite{Rosat2007}. Their
analysis shows that it is possible to improve the sensitivity by
about one order of magnitude, by applying the advanced data analysis
method. We could improve the sensitivity of the geophysical tests by
applying the advanced analysis method.

In order to make the optimum use of the global data form the GGP
network, we have to consider the optimum geometrical configuration
of the stations, which are most suitable for each test. Some of the
discussed effects exhibit latitude and/or longitude dependencies.
For example, violation signals of the universality of free-fall
vanish near the poles; observatories located near the equator are
favored for this test (see section \ref{sst:VEP}).

Another point to be considered for the optimum use of the global
data is the noise levels of the sites. The noise levels depend on
the instruments and geophysical locations \cite{Rosat2004}. We
should choose low noise sites in favored locations for each test.

By carrying out coincidence measurements with multi-station, we
could improve the sensitivity. Analysis methods for coincidence
measurements in the GGP network has to be developed.

As we have seen in sections \ref{sst:anisotopies} and
\ref{sst:GWsearch}, previous searches for the anistropies in $G$ and
gravitational waves have been done in 1960's - 1970's. Studies on
the normal modes and Earth tides have been improved significantly
since then. Also, the instruments and the global network have been
developed; the sensitivity of gravity measurements at the signal
frequencies was improved by about three orders of magnitude.
Therefore, we can expect significant improvements on the previous
results. However, as discussed earlier, there are some geophysical
effects that mimic the expected signals. It is essential to model
the unwanted effects and remove them from the data, to achieve a
good sensitivity.

The test of the universality of free-fall and search for dilatonic
waves attempt to monitor the translational motions of the inner
core. One of the targets of the GGP network is the study of
translational motions of the inner core (the Slichter triplet
\cite{Slichter1961}); geophysicists have been searching for the
Slichter triplet to determine physical properties of the Earth's
interior \cite{Slichter-search}. Therefore, the GGP network and
other technologies developed for the Slichter-triplet search can be
applied to the test of the universality and search for the dilatonic
waves. However, physical properties of the Earth's interior is not
well known. The uncertainties in the Earth model would limit these
experiments.

Fortunately, there are intensive efforts being made with new
technologies to improve the Earth model. For example, recent
advances in particle physics are providing new tools to see the
Earth's interior. Some of the examples are the detection of the
antinutrinos from natural radioactivity in the Earth with KamLAND
\cite{KamLAND2005} and studies on neutrino oscillation tomography of
the Earth's interior\cite{Winter2005}. Also, laboratory experiments
at high pressure and high temperature are being performed to
determine the viscosity of the core \cite{Mineev2004}. A new
geophysical approach of coincidence measurements with a laser
strainmeter system and a superconducting gravimeter is being carried
out at the Kamioka Observatory in Japan \cite{Takemoto2004}, to
study the normal modes, the Slichter triplet, silent earthquakes and
other geophysical phenomena. With these researches employing new
technologies, one can expect that our knowledge on the Earth model
and geophysical phenomena will be improved significantly in the near
future.

\section{Conclusions}
Superconducting gravimeters have been proved to be stable and
sensitive in geophysical studies and also they have been used to
study gravitational physics since 1970's. By using the Earth as the
test body, we have investigated possible applications of the global
network of the superconducting gravimeters to gravitational physics.

We have discussed possible improvements on the previous search for
anistropies in the gravitational constant $G$. With the GGP network
and improved knowledge on disturbing effects, we have seen that it
would be possible to achieve a comparable sensitivity with the LLR.

We have proposed to reinvestigate the normal-mode method of
searching for gravitational waves and scalar gravitational waves,
which have been attempted by pioneers of gravitational-wave research
in 1960's to 1970's, by making use of the advanced technologies in
superconducting gravimetry.

Also, we have described the proposed applications of testing the
universality of free-fall and searching for composition-dependent
dilatonic waves, using the Earth as the test body and the
superconducting gravimeters as the displacement sensor.

These geophysical tests would ultimately be limited by the
uncertainties in the Earth model. However, future improvements can
be expected from progress in new technologies and further studies
going on in geophysical studies.

\section*{Acknowledgements}
I would like to thank C. Hwang, C.-C. Chung and T. Sato for their
comments and advice on this work, and J. A. Lobo for sharing his
insights on spherical detectors for gravitational waves, and C.-W.
Lee, R. Kao and W.-C. Hsieh for the maintenance and operation of the
instruments at LOGG. This work is funded by the Ministry of Interior
of Taiwan.

%\appendix
%\section{First Appendix} %Empty argument \section{} yields `Appendix'.
%
%\section{Second Appendix}

\end{document}